\documentclass[aip,twocolumn,letterpaper,showpacs]{revtex4}

\usepackage{graphicx}
\usepackage[intlimits]{amsmath}
\usepackage{amssymb}

\newcommand{\Eq}[1]{Eq.~(\ref{#1})}

\begin{document}


\title{Non-adiabatic Josephson Dynamics in Junctions with in-Gap Quasiparticles}%



\author{J. Michelsen}
\email{jens.michelsen@chalmers.se}
\author{V.S. Shumeiko}

\affiliation{%
Department of Microtechnology and Nanoscience, MC2\\
Chalmers University of
Technology, SE-41296 Gothenburg, Sweden\\
}%


\date{\today}

\begin{abstract}

Conventional models of Josephson junction dynamics rely on the absence of low
energy quasiparticle states due to a large superconducting gap. With this
assumption the quasiparticle degrees of freedom become "frozen out" and the
phase difference becomes the only free variable, acting as a fictitious
particle in a local in time Josephson potential related to the adiabatic and
non-dissipative supercurrent across the junction. In this article we develop
a general framework to incorporate the effects of low energy quasiparticles
interacting non-adiabatically with the phase degree of freedom. Such
quasiparticle states exist generically in constriction type junctions with
high transparency channels or resonant states, as well as in junctions of
unconventional superconductors. Furthermore, recent experiments have revealed
the existence of spurious low energy in-gap states in tunnel junctions of
conventional superconductors - a system for which the adiabatic assumption
typically is assumed to hold. We show that the resonant interaction with such
low energy states rather than the Josephson potential defines nonlinear
Josephson dynamics at small amplitudes.
\end{abstract}


\maketitle 


\section{Introduction}

During the last twenty years the microscopic theory of the Josephson effect has been
undergoing steady development following the advent of novel mesoscopic
Josephson structures such as transparent metallic and semiconducting
junctions \cite{Takayanagi}, quantum point contacts \cite{Ruitenbeek},
quantum dot contacts \cite{Delft},  junctions with spin-active interfaces
\cite{Ryazanov}. Much of the theory development for these structures were
based on pioneering work by I.O. Kulik \cite{SNS,ScS,SFS,CB}. Also important
breakthrough was experimental demonstration
\cite{Nakamura1997,Saclay1998,Friedman2000} of macroscopic quantum coherence
\cite{Leggett1985} in Josephson junctions, and realization of quantum
Josephson circuits (qubits)
\cite{Makhlin2001,Wendin2007,Clark2008,Girvin2008}.

Functioning of quantum Josephson circuits is based on a fundamental property
of Josephson tunnel junctions:  nonlinear non-dissipative phase dynamics.
Equivalence of Josephson junctions to ideal nonlinear oscillators, pointed
out already by Josephson \cite{Josephson}, is used in numerous applications
 in microwave electronics \cite{Likharev}. The
possibility to quantize the motion of Josephson  oscillator \cite{Scalapino},
and to observe the macroscopic quantum dynamics is essentially based on this
fundamental property.

Equation of motion for the superconducting phase difference across the
junction stems from Kirchhoff's rule that combines the Josephson tunneling
current, $I_J(\varphi)=I_C\sin\varphi$, and the displacement current through
junction capacitor, $( C/2e) \ddot\varphi$,
\begin{equation}\label{classic}
{ C\over 2e}\,\ddot\varphi + I_C\sin\varphi = I_e(\varphi,t),
\end{equation}
where $I_e(\varphi,t)$  is a biasing current defined by external circuit,
 $\hbar=1$. A key assumption behind this
equation is a quasi-static form of the Josephson current that extends the
static current-phase relation to the non-stationary case of temporal
variation of the phase. A justification for this assumption is provided by a
wide isotropic superconducting energy gap $\Delta$ that prevents excitation
of quasiparticles by temporal variation of the phase at low temperature and
small frequency of Josephson plasma oscillation, $kT,\hbar\omega_p\ll\Delta$.
Thus electrons in the junction remain in equilibrium, and the adiabatic form
of the Josephson current is maintained.

Such an approach is relevant for tunnel junctions, but it is not always
correct. Notable exceptions are transparent point contacts
\cite{Beenakker1991} and resonant quantum dot contacts \cite{Super96}
containing Andreev bound states deep inside the energy gap. Other important
exceptions are junctions of d-wave superconductors containing zero energy
Andreev surface states \cite{Lofwander2001} and low energy nodal
quasiparticles \cite{Scalapino1995}. In such junctions the low energy
quasiparticles are involved in the  macroscopic dynamics: they are excited
and driven away from equilibrium by temporal variation of the phase resulting
in significant modification of the Josephson current. How is \Eq{classic}
then modified in the presence of low energy quasiparticle states?

In this article we suggest an extension of \Eq{classic} to describe the
non-adiabatic Josephson dynamics in the presence of interaction with
quasiparticles. A general equation derived in the next sections has the
form,
\begin{equation}\label{nonad}
\begin{split}
\frac{ C}{2e}\,\ddot{\varphi} + \mathrm {Tr}\,(\hat I_J \hat f)=I_e,
\\
i \dot{\hat{f}}=[\hat H\,,\,\hat{f}].
\end{split}
\end{equation}
Here the adiabatic Josephson current is replaced by a statistical average of
a Josephson current operator, $\hat I_J$; the non-equilibrium quasiparticle
density matrix $\hat f$ satisfies the Liouville equation with an effective
Hamiltonian, $\hat H$. The only approximation made during the derivation is a
semiclassical approximation for the phase dynamics, otherwise this is an {\em
exact equation}. As we will show, both the current operator and effective
Hamiltonian are expressed through the quasiparticle energy spectrum of the
junction and interlevel transition matrix elements.

\Eq{nonad} has a generic form of equation of motion of a macroscopic particle
interacting with a fermionic bath. Usually such problems are treated assuming
an equilibrium bath. Here we will consider a non-equilibrium bath consisting
of low energy bound Andreev states strongly driven by the phase dynamics. Our
main conclusion is that the Rabi dynamics of the Andreev states dramatically
modifies the nonlinear properties of macroscopic Josephson dynamics. The
physics here resembles well known in nonlinear optics picture of interaction
of electromagnetic mode with medium of two-level atoms \cite{Eberly}.

The structure of the paper is  as follows. In section II we discuss a general
approach based on the path integral technique, which is used in Section III
to derive \Eq{nonad}. In the next section we discuss the adiabatic limit and
establish connection between our method and earlier results for tunnel
junctions. Section V is devoted to non-adiabatic effects; we study both the
linear and nonlinear quasiparticle response, the main result here is the
evaluation of a nonlinear effect of driven low energy Andreev bound states.
In Section VI we present the derivation of stochastic Langevin equation
generalizing the deterministic \Eq{nonad}.


\section{Formulation of the problem}
Consider a general setup of a junction with  superconducting electrodes
occupying left ($x<0$) and right ($x>0$) halfspaces, with an interface at
$x=0$  carrying $N$ conducting modes. We will not specify the properties
of the interface but rather characterize it, within the quasiclassical
approximation, with some electronic transfer matrix. In the following we
also adopt common assumptions: (i) the superconductors are described with
a BCS mean field theory, (ii) superconducting electrodes maintain local
equilibrium implying absence of spatial and temporal variation of the
module and phase of the order parameter, $\Delta=\mathrm{const}$,
$\chi(\mathbf{r},t)=\text{sign}(x)\varphi(t)/2$.

To accurately describe the  nonequilibrium dynamics we adopt the path
integral approach, introduced by Ambegaokar et al. \cite{AES1}, and adapted
for non-equilibrium systems \cite{Kamenev2005,Kleinert,Weiss}. Following this
approach we represent the trace of the time dependent statistical operator of
the junction $\hat{\rho}(t)=\hat{U}(t,t_0)\hat{\rho}_0\hat{U}(t_0,t)$ with
the path integral,
\begin{equation}\label{Z}
Z=\text{Tr}[\hat{\rho}(t)]=\int \mathcal{D}\varphi
\mathcal{D}\bar{\psi}\mathcal{D}\psi e^{iS[\varphi,\bar{\psi},\psi]},
\end{equation}
where the action is,
\begin{equation}\label{action}
S[\varphi,\bar{\psi},\psi]=\int_\mathcal{C}dt
\left(\frac{C}{8e^2}\dot{\varphi}^2-U_e(\varphi)+(\bar{\psi},
\mathcal{G}^{-1}\psi)\right).
\end{equation}
The first term in this equation originates from the electrostatic interaction
between electrodes and is described within the capacitance approximation
\cite{AES1};  the second term is an inductive energy of the external circuit,
and the last term represents the contribution of superconducting electrons.
Time integration  goes along the forward-backward time contour,
$\mathcal{C}=\mathcal{C}_++\mathcal{C}_-$. The fermionic fields in the
electronic term are written in the Nambu pseudo-spinor representation,
$\psi=(\psi_\uparrow , \bar{\psi}_\downarrow)^T$, and
\begin{equation}\label{G-1}
\mathcal{G}^{-1} =i\partial_t-\mathcal{H} (\varphi) -{\dot{\chi}\over
2}\,\sigma_z,
\end{equation}
where
\begin{equation}\label{bdgham}
\mathcal{H}(\varphi)=\left({\boldsymbol{p}^2\over 2m} -E_F+
V(\mathbf{r})\right)\sigma_z + \Delta e^{ i\sigma_z \chi}\,\sigma_x,
\end{equation}
is the junction Hamiltonian. Here $V(\mathbf{r})$ is the potential
defining the interface; superconducting order parameter, $\Delta$, is a
scalar in s-wave superconductors, but becomes a nonlocal operator in the
case of unconventional d-wave pairing. The last term in \Eq{G-1}
represents the electrical potential needed to preserve electro-neutrality
within the electrodes \cite{ZAZ2},

By virtue of the quadratic form of the Fermionic part of the action
(\ref{action}) one can formally perform the gaussian path integral over the
fermionic fields, and reduce the integral to one over the phase degree of
freedom \cite{AES2,LO,ZP,ZS},
\begin{equation}\label{effectiveactionG}
Z= \int \mathcal{D}\varphi \ e^{\displaystyle iS_0[\varphi]+\text{Sp}\ln
(-i{\cal G}^{-1})},
\end{equation}
here $S_0$ comprises the first  two terms in \Eq{action}, and $\text{Sp}$
denotes the trace over both the quasiparticle states as well as the
forward-backward time contour. This transformation in itself, however, does
not solve the problem: the obtained effective action contains the contour
ordered fermionic Green's function, which needs to be computed by solving the
equation of motion. This can only be done under some approximations. The most
studied in literature case  concerns tunnel junctions where the Green's
function is calculated perturbatively using small transparency of the
junction, $D\ll 1$ \cite{AES1,AES2,LO}. This is commonly done within the
formalism of tunnel Hamiltonian model. This method can be improved and made
suitable for transparent junctions, $D\sim 1$, by performing summation of the
whole perturbative series \cite{CUEV}. However, the tunnel model method does
not straightforwardly apply to superconductors with surface states, such as
d-wave superconductors, since it is based on expansion over bulk Green's
functions. The tunnel model must then be modified by considering
semi-infinite leads with hard-wall boundaries rather than homogenous leads
\cite{FOG}. An alternative way to calculate the effective action for transparent
junctions was suggested in Refs. \cite{ZAZ,ZAZ2}, by using exact boundary
conditions and an adiabatic approximation for low energy Andreev states.
Zaikin and Panuykov \cite{ZP,ZS} suggested a general method for calculating
the effective action by establishing a formal relation between the action and
the current across the junction. This method, however, requires knowledge of
the ac current response to an arbitrary time dependent realization of
$\varphi(t)$, which in general is not possible to obtain.

In this paper we suggest an  alternative method of calculation of the
effective action (\ref{effectiveactionG}), which is {\em exact} in the limit
of semiclassical phase dynamics, and {\it universal} regarding interaction
with any kind of quasiparticle states.

\subsection{Instantaneous Basis}

The central idea of the method is to expand the Nambu fields over an {\it
instantaneous}  eigenbasis of the Hamiltonian (\ref{bdgham}),
\begin{equation}\label{}
\psi(\mathbf{r},t)=\sum_i\phi_i(\mathbf{r};\varphi) a_i(t).
\end{equation}
This allows us to separate the spatial problem from the temporal one by
solving the time independent Bogoliubov-de Gennes equation for a fixed value
of the phase. Apart from the technical simplifications this basis provides an
intuitive understanding of the microscopic processes involved in the
Josephson dynamics in terms of transitions between quasiparticle states.

In this basis the action (\ref{action}) becomes,
\begin{equation}\label{instaction}
S[\varphi,\{\bar{a}_i\},\{a_i\}]\hspace{-3pt}=
\hspace{-5pt}\int_\mathcal{C}dt\left(\frac{C}{8e^2}
\dot{\varphi}^2\hspace{-2pt}-U_e(\varphi)\hspace{-2pt}+
\hspace{-2pt}\sum_{ij}\bar{a}_iG^{-1}_{ij}a_j\right),
\end{equation}
where
\begin{equation}\label{Gij}
G_{ij}^{-1}=\left(i\partial_t-H_{ij}\right)
\end{equation}
represents the quasiparticle Green function in the instantaneous basis,
and the Hamiltonian is given by equation,
\begin{equation}
H_{ij}(\varphi,\dot{\varphi})=E_i(\varphi)\delta_{ij}-
\dot{\varphi}\mathcal{A}_{ij}.
\end{equation}
The diagonal elements here are given by the instantaneous eigen energies
of the Hamiltonian (\ref{bdgham}),
\begin{equation}\label{}
\mathcal{H}\phi_i=E_i\phi_i,
\end{equation}
and the off-diagonal elements are proportional to the matrix elements,
\begin{equation}
\mathcal{A}_{ij}=\left(\phi_i, i\partial_\varphi\phi_j\right)
-\tfrac{1}{4}\left(\phi_i,\text{sign}(x)\sigma_z\phi_j\right),
\end{equation}
of the transitions between the instantaneous eigenstates due to temporal
variations of the phase.

The physical meaning of the transition matrix elements can be understood by
establishing their connection to the Josephson current operator. Consider a
general quantum mechanical equation for the charge current density matrix,
\begin{equation}
\mathbf{j}_{ij}(\mathbf{r})=\frac{ie}{2m}
\left.(\mathbf{\nabla}-\mathbf{\nabla}')\phi_i^\dag(\mathbf{r})\phi_j
(\mathbf{r}')\right|_{\mathbf{r}=\mathbf{r'}}.
\end{equation}
The current through  the interface, $S$, is given by equation,
\[I_{ij}=\int_S d\mathbf{n}\cdot \mathbf{j}_{ij}(\mathbf{r}).\]
This is the matrix of the Josephson current operator. If we connect the
electrodes in a loop at infinity, we can use the fact that no current is
flowing through any other part of the surface of the superconductor so we may
extend the surface, $S$, around the whole superconductor and use Gauss law:
\begin{equation}
2I_{ij}=\int_{L} d^3r \  \mathbf{\nabla}\cdot \mathbf{j}_{ij}
(\mathbf{r})-\int_{R} d^3r \  \mathbf{\nabla}\cdot
\mathbf{j}_{ij}(\mathbf{r}).
\end{equation}
From the explicit form of the Hamiltonian (\ref{bdgham}) we derive the
relation,
\begin{equation}\begin{split}
(-i\mathbf{\nabla})\cdot\mathbf{j}_{ij}&=-\frac{e}{2m}
\left[[-\nabla^2\phi_i]^\dag\phi_j-\phi_i^\dag[-\nabla^2\phi_j]\right]\\
&=-e\left[(E_i-E_j)\phi_i^\dag\sigma_z\phi_j+\phi^\dag_i[\mathcal{H},
\sigma_z]\phi_j\right].\\
\end{split}
\end{equation}
The last term  in this equation can be rewritten as
\begin{equation}
[\sigma_z,\mathcal{H}]=4 i\text{sign}(x)\,\partial_{\varphi} \mathcal{H}.
\end{equation}
The current operator then becomes
\begin{equation}
\begin{split}
&I_{ij}=\\
=&-2ie\left[(\phi_i, i\partial_\varphi\mathcal{H}\phi_j -
\frac{1}{4}(E_j-E_i) (\phi_i,\text{sign}(x)\sigma_z\phi_j)\right].
\end{split}
\end{equation}
By differentiating the eigenvalue  equation, $\mathcal{H}\phi_i=E_i\phi_i$,
with respect to $\varphi$ one obtains the following identities,
\begin{equation}\begin{split}
&(\phi_i,i\partial_\varphi\mathcal{H}\phi_i)=i\partial_\varphi E_i,\\
&(\phi_i,i\partial_\varphi\mathcal{H}\phi_j)=(E_j-E_i)(\phi_i,
i\partial_\varphi\phi_j), \quad i\neq j.
\end{split}.
\end{equation}
From these one sees that the current matrix elements are given by
equations,
\begin{equation}\begin{split}
I_{ii}&=2e\partial_\varphi E_i,\\
I_{ij}&=2e i(E_i-E_j)\mathcal{A}_{ij},
\end{split}
\end{equation}
or
\begin{equation}\label{Iij}
I_{ij}=2e\left(\frac{\partial E_i}{\partial\varphi}\,\delta_{ij} +
i[E,\mathcal{A}]_{ij}\right).
\end{equation}
Thus we conclude that the matrix elements $\mathcal{A}_{ij}$ are related to
the off-diagonal matrix elements of the Josephson current operator.

Towards the end of this section we present a many body Hamiltonian of the
junction in the instantaneous eigen basis. To this end we define the
conjugate momentum $n$ corresponding to $\varphi$,
\begin{equation}
n=\frac{\partial L}{\partial \dot{\varphi}}=
\frac{C}{4e^2}\,\dot{\varphi}+\sum_{ij}\mathcal{A}_{ij}\bar{a}_ia_j,
\end{equation}
and perform a Legendre transformation of the Lagrangian  in Eq.
(\ref{instaction}), then we promote the variables, $\bar{a}_i,a_i$, and
$\varphi, n$ to operators by imposing standard (anti-) commutation relations
to get,
\begin{equation}\begin{split}
\mathcal{H}_q &= {(2e)^2\over 2C}\left(\hat{n}+
\sum_{ij}\mathcal{A}_{ij}(\varphi)\hat{a}_i^\dag\hat{a}_j\right)^2\\
&+U_e(\varphi)+\sum_i E_i(\varphi)\hat{a}_i^\dag \hat{a}_i.
\end{split}
\end{equation}
%

\section{Equation of motion}

Now we perform integration over the fermionic variables using the
instantaneous eigen basis,
\begin{equation}\label{effectiveaction}
Z =\int \mathcal{D}\varphi \ e^{iS_0[\varphi]+\text{Sp}\ln (-iG^{-1})} =
\int \mathcal{D}\varphi \ e^{iS_\text{eff}[\varphi]}.
\end{equation}
Defining in a standard manner four Green's function components, depending on
wether the time arguments are defined on the forward ($a=+$) or backward
($a=-$) part of the contour,
\begin{equation}
G(t,t')=G^{ab}(t,t'), \quad t\in \mathcal{C}^a, \ t'\in \mathcal{C}^b,
\end{equation}
we write \Eq{Gij} on the form,
\begin{equation}\label{gfequation}
a\left(i\partial_t - \hat{H}(\varphi^a,\dot{\varphi}^a)\right)
\hat{G}^{ab}(t,t')=\delta^{ab}\delta(t-t').
\end{equation}
Introducing a single particle density matrix through the relation,
\begin{equation}\label{}
\hat{f}(t)=\frac{1}{2i}\sum_{a}\hat{G}^{aa}(t,t),
\end{equation}
we get from \Eq{gfequation} the Liouville equation,
\begin{equation}\label{Liouville}
i\dot{\hat{f}}=[\hat{H},\hat{f}], \quad \hat H=\hat E -
\dot\varphi\hat{\cal A}.
\end{equation}

A semiclassical dynamical equation for the superconducting phase is given
by the least action principle formulated in terms of the Wigner variables,
$\varphi^a=\varphi+a \chi/ 2$, and has the form \cite{Kamenev2005},
\begin{equation}\label{eqphi}
\left.{\delta S_\text{eff}[\varphi,\chi]\over\delta \chi}\right|_{\chi=0}=0.
\end{equation}
To  calculate the functional derivative of the fermionic part, we perform
a rotation to a single particle basis, in which the dependence on the time
derivative of the phase is eliminated from the Hamiltonian. This is
achieved by using a unitary matrix $ \hat{U}(\varphi)$ satisfying the equation
$i\partial_\varphi\hat{U}=-\hat{\mathcal{A}}\hat{U}$. Computing the
derivative and rotating back to the original basis we find,
\begin{equation}\label{avgcurrent}
\left.\delta \text{Sp}\ln[-i\check{G}^{-1}] \over \delta\chi\right| _{\chi=0}
= {i\over 2e}\, \text{Tr}\left(\hat{I}_J(\varphi)\hat{f} \right)\equiv
{i\over 2e}\langle \hat{I}_J\rangle,
\end{equation}
where $\hat I_J(\varphi)$ is the Josephson current operator defined in
\Eq{Iij}. Then introducing external current, $I_e=- 2e\partial_\varphi U_e$,
we write equation of motion on the form,
\begin{equation}\label{classicaleqm}
\frac{C}{2e}\,\ddot{\varphi} + \text{Tr}(\hat{I}_J\hat{f})= I_e, \quad
\hat I_J =2e\left(\partial_\varphi \hat E + i[\hat
E,\hat{\mathcal{A}}]\right).
\end{equation}
Eqs. (\ref{Liouville}) and (\ref{classicaleqm})  together constitute a
central technical result of this paper.

\section{Adiabatic Limit}

In general, in order to solve the coupled equations for the phase
(\ref{classicaleqm}) and the density matrix (\ref{Liouville}), one needs to
calculate  the static quasiparticle energy spectrum, and matrix elements of
the interlevel transitions. This is a rather difficult task since the latter
quantities are complicated functions of the phase. However, if the
quasiparticle spectrum has a gap, and the frequency of the plasma oscillation
is small compared to this gap, in other words, if the quasiparticle dynamics
is fast on the time scale of the phase variation, one can apply an adiabatic
approximation to find the solution.

A formal condition for the adiabatic expansion is
$\dot{\varphi}\mathcal{A}_{ij}\ll E_i-E_j$. In the main approximation, the
Hamiltonian in \Eq{Liouville} reads, $\hat H = \hat E$, and the initial
equilibrium density matrix, $\hat f(0)=  f^0(\hat E(0))$ defines the solution
that remains constant during the phase evolution, $\hat f(t) = \hat f^0$.
This implies that the trace in \Eq{classicaleqm} will only contain  the
diagonal part of the current operator, and the Josephson current reduces to
the adiabatic form,
\begin{equation}\label{adiabaticI}
\begin{split}
I_J^\text{ad}(\varphi)= 2e\text{Tr}(\partial_\varphi\hat E\hat{f}^0)=
2e\partial_\varphi U_J,\\ U_J(\varphi) = \text{Tr}(\hat
E(\varphi)\hat{f}^0).
\end{split}
\end{equation}
This equation provides a generalization of the tunnel junction equation
(\ref{classic}) to the junctions with non-sinusoidal current-phase
dependence.

To find the first non-adiabatic correction, it is convenient to expand
electronic part in effective action, \Eq{effectiveaction},
\begin{equation}\label{series}
\begin{split}
\text{Sp}\ln(-iG^{-1})&=  \text{Sp}\ln(-i(G^\text{ad})^{-1})\\
&-\sum_{n}\frac{1}{n}\, \text{Sp}\left(-\check{G}^\text{ad} \dot\varphi \check
{\mathcal A}\right)^{ n},
\end{split}
\end{equation}
where $(G^\text{ad})^{-1}=\delta_{ij}(i\partial_t-E_i(\varphi))$.  The first,
adiabatic term is given by equation,
\begin{equation}
\text{Sp}\ln\left(-i[G^\text{ad}]^{-1}\right)=-i\int_\mathcal{C} dt \
U_J(\varphi),
\end{equation}
consistent with \Eq{adiabaticI}. To see this, we formally introduce
$G_\lambda^\text{ad}=G^\text{ad}(\lambda\varphi)$, and rewrite the adiabatic
term as \cite{ZS},
\begin{equation}\begin{split}
&\text{Sp}\ln\left(-i[G^\text{ad}]^{-1}\right) = \int^1
d\lambda\frac{d}{d\lambda}\text{Sp}\ln\left(-i(G_\lambda^\text{ad})^{-1}
\right)\\
&= -\int_\mathcal{C} dt \int^1 d\lambda  \frac{\partial
E_i(\lambda\varphi)}{\partial\lambda} [G^\text{ad}_\lambda]_{ii}(t,t).
\end{split}
\end{equation}
Since $[G^\text{ad}_\lambda]_{ii}(t,t)=if_{i}^0$ does not change with time by
virtue of the earlier presented argument, we find,
\begin{equation}\begin{split}
&\text{Sp}\ln\left(-i[G^\text{ad}]^{-1}\right)  \\
&= -i\int_\mathcal{C} dt \int^\varphi d\varphi'  \text{Tr}
\left(\frac{\partial \hat E(\varphi')}{\partial\varphi'} \hat f^0 \right)=
-i\int_\mathcal{C} dt U_J.
\end{split}
\end{equation}
The first order non-adiabatic term in  the series, \Eq{series}, cancels since
$\hat{G}^\text{ad}$ is diagonal while $\hat{\mathcal A}$ is purely
off-diagonal, which implies that the trace of their product is zero. Keeping
then only the second order correction we find,
%
\begin{equation}\begin{split}
&\text{Sp}\ln(-i\check{G}^{-1}) =-i\int_\mathcal{C} dt \ U_J(\varphi)\\
&-\frac{1}{2}\int_\mathcal{C}dt \int_\mathcal{C}dt'
\dot{\varphi}\mathcal{A}_{ij}(\varphi)
G_{jj}^\text{ad}(t,t')\dot{\varphi}'\mathcal{A}_{ji}(\varphi')
G^\text{ad}_{ii}(t',t).
\end{split}
\end{equation}
%
When the occupied and unoccupied states are separated  by a large gap, the
product
\begin{equation}\label{}
G_{jj}^\text{ad}(t,t')G_{ii}^\text{ad}(t',t)\sim e^{-i\int_{t}^{t'}
dt''\varepsilon_{ij}(\varphi'')}, \quad \varepsilon_{ij}=E_i-E_j,
\end{equation}
oscillates rapidly on the scale of variations of the phase, and we can treat this object
in the local approximation. This gives us,
\begin{equation}
\text{Sp}\ln(-i\check{G}^{-1})=-i\int_\mathcal{C} dt \left[\ U_J(\varphi)
+ \frac{\delta
C(\varphi)}{8e^2}\,\dot{\varphi}^2\right],
\end{equation}
where
\begin{equation}\label{}\label{deltaC}
\delta C(\varphi)=2e^2\sum_{ij}
\frac{|\mathcal{A}_{ij}(\varphi)|^2f_i(1-f_j)}{\varepsilon_{ij}(\varphi)}
\end{equation}
represents a phase dependent correction to the junction capacitance.

Let us  explicitly evaluate the contribution to \Eq{deltaC} of the Andreev
bound states in a tunnel junction. In tunnel junctions, Andreev energy levels
are located very close to the gap edges \cite{Furusaki} having the level
spacing, $\varepsilon=2\Delta\sqrt{1-D\sin^2\tfrac{\varphi}{2}}\approx
2\Delta$. The transitions connect only Andreev states of the same conducting
mode with transition matrix elements \cite{PRB93},
\begin{equation}\label{}
\mathcal{A}=i\sqrt{RD}\Delta{|\sin\varphi/2|\over 2\varepsilon}\approx
{i\sqrt{D}\over4}\sin\frac{\varphi}{2}.
\end{equation}
Computing the correction to the capacitance using these expressions we find
the phase dependent correction in the zero temperature limit to be, $\delta
C_\varphi\approx (De^2/32\Delta)\cos\varphi$, per conducting mode. This is consistent
with the result of the tunnel model calculation in Refs. \cite{AES1,AES2}.

\section{Non-adiabatic dynamics}

\subsection{Linear response}
The non-adiabatic dynamics essentially results from the {\em resonant}
response of low energy quasiparticles to the phase variation. In this section
we consider the linear quasiparticle response and compute the non-adiabatic
correction to the frequency of Josephson plasma oscillation.

Consider small deviations from an equilibrium configuration,
$\varphi=\varphi_0$ and $\hat{f}=\hat{f}^0$ determined by the equation,
$I_J^\text{ad}(\varphi_0)=\text{Tr}(\hat{I}(\varphi_0)
\hat{f}^0)=I_e(\varphi_0)$. Straightforward linearization of Eqs.
(\ref{Liouville}) and (\ref{classicaleqm}) with respect to small deviations
of the phase, $\varphi(t)-\varphi_0$, and the density matrix, $\hat f(t)-\hat
f^0$, leads to the dispersion equation for the plasma oscillation,
\begin{equation}\label{dispeq}
\left(-\omega^2+\omega_p^2+\omega\gamma_0(\omega)\right)\varphi_\omega=0,
\end{equation}
where
\begin{equation}\label{}
\omega_p^2=\frac{2e}{C}\,\frac{\partial
I_J^\text{ad}(\varphi_0)}{\partial\varphi}
\end{equation}
is the adiabatic plasma frequency, and $\gamma_0(\omega)$ denotes the linear
response of the quasiparticles,
\begin{equation}\label{linearresponse}
\gamma_0(\omega)={4e^2\over C}\,
\sum_{ij}\frac{\varepsilon_{ij}|\mathcal{A}_{ij}|^2
(f_{i}^0-f_{j}^0)}{\varepsilon_{ij}-(\omega+i0)}.
\end{equation}
The linear response  of quasiparticle is a relevant approximation at small
phase oscillation when the quasiparticles have a continuous energy spectrum
and the transferred energy is dispersed across a large phase space volume
resulting in weak non-equilibrium. As such the dispersion equation
(\ref{dispeq}) can be applied, for example, to the low energy itinerant
states  in the nodal regions of high-Tc superconductors, or to broadened
Andreev bound states in disordered junctions. However, the linear
approximation does not apply to spectroscopically narrow Andreev bound
states, whose response is essentially nonlinear even at small phase
amplitude.

\subsection{Resonant interaction with Andreev levels}
Now we consider the nonlinear  dynamics of the phase  driven by small
oscillating current $I_e(t)=I_e\cos\omega t$, at a frequency not far from the
resonant frequency, $\delta=\omega-\omega_p\ll 1$, in the presence of
resonant interaction with weakly broadened low energy Andreev levels. Such
levels may exist in transparent electronic conducting modes close to
$\varphi_0=\pi$, in electronic modes with resonant transmissivity, or in
surface modes of d-wave superconductors. The exact nature of these states
does not play any role for our analysis. The important properties are: (i)
the phase variations do not change the electronic momentum hence do not
induce quasiparticle transitions among the conducting modes, (ii) therefore
transitions only occur between pairs of Andreev states within the same
conducting mode, (iii) the Andreev levels are well separated from the
continuum states of the mode. Under these assumptions, the Hamiltonian in
\Eq{Liouville} truncated to the Andreev level subspace consists of a sum of
independent two-level Hamiltonians, and the density matrix factorizes to the
product of two-level density matrices parameterized with the conduction mode
number, $\hat f(n)$. The non-adiabatic current then becomes:
\begin{equation}\begin{split}
\langle &I_J\rangle -I_J^\text{ad}(\varphi)=\\
&2e\sum_{n}\left(\frac{\partial\varepsilon}{\partial\varphi}(f_z-f_{z}^0)
-\varepsilon(\mathcal{A}f_- +\mathcal{A}^*f_+ )\right),
\end{split}
\end{equation}
where,  $\varepsilon=E_1-E_2$, is the level spacing between two Andreev
states associated with a specific mode $n$ and,
$i\mathcal{A}=\mathcal{A}_{12}$, is the corresponding transition matrix
element (we skip index $n$ for brevity). Similarly  $f_{z}=f_{11}-f_{22}$ and
$f_+=f_{12}=(f_-)^*$, are the corresponding elements of the two-level density
matrix satisfying the Bloch-Redfield equation,
\begin{equation}\begin{split}
\dot{f}_+&=(-i\varepsilon - \Gamma_2) f_++2\dot{\varphi}\mathcal{A} f_z\\
\dot{f}_z&=-\dot{\varphi}\mathcal{A}f_+^{\ast}-\dot{\varphi}\mathcal{A}^\ast
f_+  -\Gamma_1(f_z-f_{z,0}),
\end{split}
\end{equation}
where we have  added  phenomenological decay rates $\Gamma_1$ and $\Gamma_2$
originating, e.g., from  some weak inelastic interaction with the continuum
states.

To separate the fast and slow resonant dynamics, we parameterize the phase
as,
\begin{equation}\begin{split}
\varphi(t)&=\frac{1}{2}(\varphi_\omega(t) e^{-i\omega t}+c.c.)\\
\dot{\varphi}(t) &=\frac{\omega}{2i}(\varphi_\omega(t) e^{-i\omega t}-c.c.),
\end{split}
\end{equation}
where the complex variable,  $\varphi_\omega(t)=r(t)e^{i\vartheta(t)}$,
depends on the amplitude of oscillations, $r(t)$, and the time dependent
phase shift, $\vartheta(t)$.  Using a similar separation for the fast and
slow parts of the off-diagonal elements of the density matrix,
\begin{equation}
f_+(t)=f_\omega(t) e^{-i\omega t},
\end{equation}
we get, after expanding to first order in,  $\varphi-\varphi_0$, and
averaging over fast variables (note $\mathcal{A}_0=\mathcal{A}(\varphi_0)$
and $\varepsilon_0=\varepsilon(\varphi_0)$),
\begin{equation}\label{slowfeq}
\begin{split}
\dot{f}_\omega&=-i(\varepsilon_0 -\omega-i\Gamma_2)
f_\omega-i\omega\mathcal{A}_0
\varphi_\omega f_z\\
\dot{f}_z&=i\frac{\omega}{2}(\mathcal{A}_0\varphi_\omega
f^{\ast}_\omega+c.c.)-\Gamma_1(f_z-f_{z,0}).
\end{split}
\end{equation}
The regime relevant for our discussion corresponds to slow variation of the
phase oscillation envelope, $\varphi_\omega$, on the  time scale of the
Andreev state relaxation. Then the Andreev state density matrix will
adiabatically follow the evolution of the phase amplitude (in the rotating
frame), and we restrict ourselves to the quasi-stationary solutions,
$\dot{f}_\omega,\dot{f}_z\approx 0$, to find from the first equation in
(\ref{slowfeq}),
\begin{equation}
f_\omega= \frac{\omega\mathcal{A}_0\varphi_\omega}{(\omega+i\Gamma_2)
-\varepsilon_0}f_z.
\end{equation}
Inserting this expression into the current we find,
%
\begin{equation}\label{currentresponse}
\begin{split}
\langle I_J\rangle &=I_J^\text{ad}
(\varphi)+2e\sum_n\left(\frac{\partial\varepsilon_0}{\partial\varphi}
(f_z-f_{z}^0) \right.\\
& \left. -
\frac{\omega\varepsilon_0|\mathcal{A}_0|^2f_{z}}{(\omega+i\Gamma_2)
-\varepsilon_0} \,\varphi_\omega e^{-i\omega t}+c.c. \right).
\end{split}
\end{equation}
%
%
Eq. (\ref{currentresponse})  illustrates the principal effect of the
resonant interaction between the phase and the Andreev levels: the phase
oscillation drives the Andreev levels to a nonequilibrium state determined
by the stationarity condition,
\begin{equation}\begin{split}
f_z&= f_{z}^0-\frac{\Omega^2(r)(\Gamma_2/\Gamma_1)}
{(\varepsilon_0-\omega)^2+\Gamma_2^2+\Omega^2(r)
(\Gamma_2/\Gamma_1)}\,f_{z}^0.
  \end{split}
\end{equation}
Here $\Omega(r)=|\omega\mathcal{A}_0r|$ is the amplitude  dependent Rabi
frequency of the Andreev two-level system associated with specific mode $n$.
The first term inside the bracket in Eq. (\ref{currentresponse}) produces a
{\it nonlinear modulation} of the Josephson potential due the nonequilibrium
population of the Andreev levels. The second term causes a {\em nonlinear
damping} of the phase oscillation,  similar to the imaginary part of the
linear response, although it now depends on the nonequilibrium population of
the Andreev levels.

For the levels  close to the resonance, $\varepsilon_0\approx \omega$, the
diagonal elements are approximately given by
\begin{equation}
f_z\approx \frac{\Gamma^2}{\Gamma^2+\Omega^2(r)} \,f_{z}^0,
\end{equation}
where, $\Gamma=\sqrt{\Gamma_1\Gamma_2}$.  Thus in  the limit of $\Omega(r)\ll
\Gamma$, i.e. $r\ll \Gamma/|\omega\mathcal{A}_0|$, we recover the linear
response regime. In the opposite limit, $\Omega(r)\gg \Gamma$, i.e. $r\gg
\Gamma/|\omega\mathcal{A}_0|$, the levels become saturated, $f_z\approx0$,
and can no longer absorb energy from the phase oscillation, thus the damping
decreases for large amplitude of phase oscillation.

\subsection{Nonlinear phase dynamics}

\begin{figure}[h!]
\begin{center}
\includegraphics[width=0.4\textwidth]{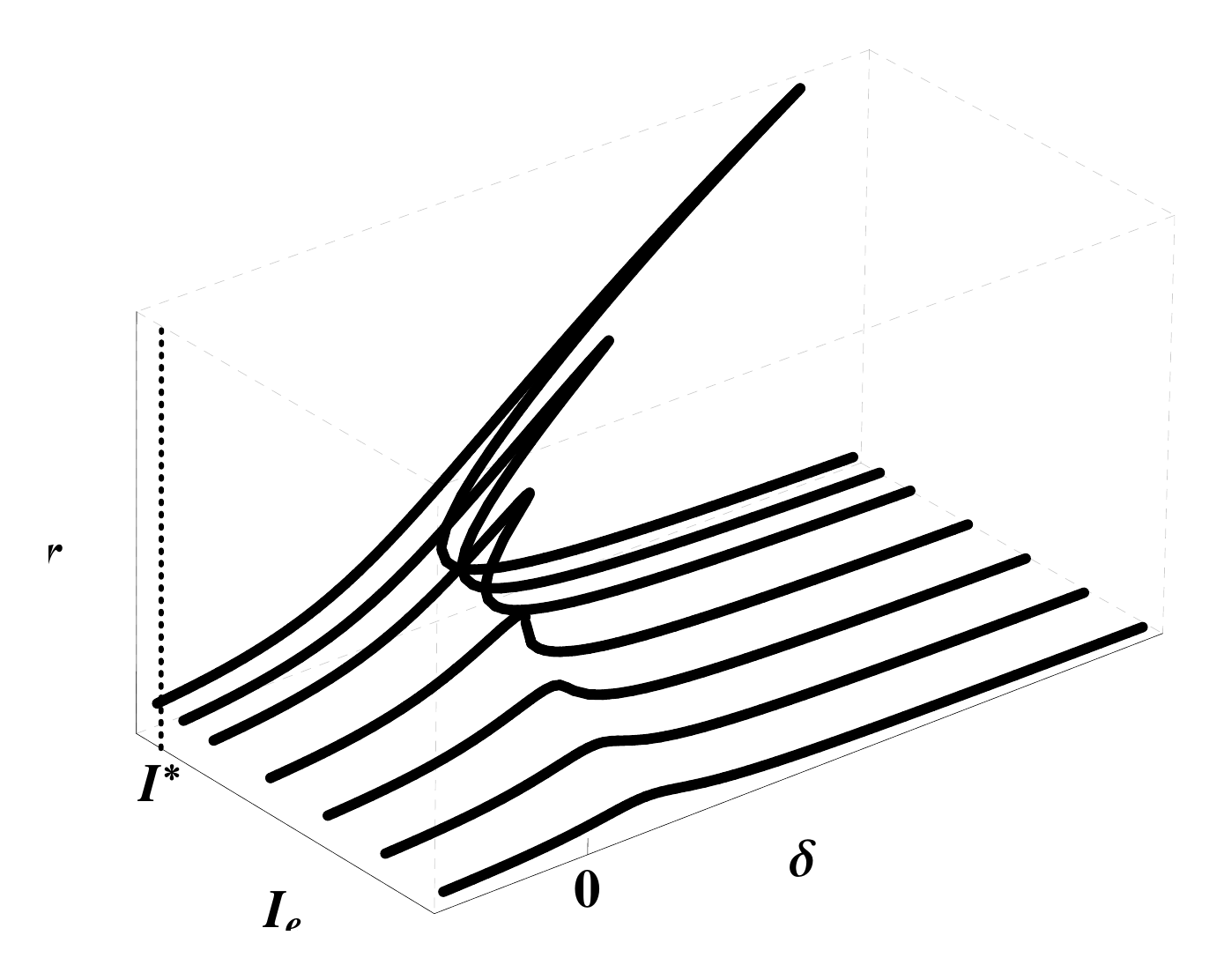}
\caption{Effect of resonant interaction with spectroscopically sharp Andreev
bound states on non-linear response of the junction. Phase oscillation
amplitude as  a
 function of detuning shown for different amplitudes of driving current.}
\label{responsefig}
\end{center}
\end{figure}

To see how the nonlinear quasiparticle response manifests itself in  the
junction dynamics we write down the equation of motion for the slowly varying
amplitudes, $\varphi_\omega$, and introduce a nonlinear response function,
$\gamma(r)=\gamma'(r)+i\gamma''(r)$, defined through the relation,
\begin{equation}\label{}
\frac{2e}{C}\left(\langle I_J\rangle-I_J^\text{ad}(\varphi)\right)= \omega
\gamma(\omega,r) \varphi_\omega e^{-i\omega t}+ c.c.,
\end{equation}
in terms of which the averaged equation for the envelope becomes,
\begin{equation}\label{slowdynamics}
-2i\omega_p\dot{\varphi}_\omega +\left[-2\omega_p\delta
+\omega_p\gamma(\omega_p,r)\right]\varphi_\omega =\frac{e}{C}{I_e}.
\end{equation}
The stationary solutions to this equation,  $\dot{\varphi}_\omega=0$, connect
resonant amplitude and detuning $\delta$,
\begin{equation}
\delta=\frac{1}{2}\gamma'(r)\pm
\frac{1}{2r}\sqrt{(e{I}_e/C\omega_p)^2-(\gamma''(r))^2 r^2}.
\end{equation}
The two solutions correspond to the stable/unstable  branches of the function
$r(\delta)$ as illustrated on Fig. \ref{responsefig}. The maximum response,
$r_m$, is found where the two branches coincide, i.e.
$r_m\gamma''(r_m)=eI_e/C\omega_p$.

To make a quantitative analysis we write $\sum_n=\int d\varepsilon
\nu(\varepsilon)$, where $\nu(\varepsilon)=\sum_n\delta
(\varepsilon-\varepsilon_0(n))$.
If the density of states $\nu(\varepsilon)$ is a smooth  function close to
the resonance the integration can be explicitly performed, giving,
\begin{equation}\begin{split}
\gamma'(r) &= \gamma_0'-
\frac{\partial_\varphi^2\bar{\varepsilon}_0r^2}{\Gamma_1}
\frac{\Gamma\gamma_0''}{\sqrt{\bar{\Omega}(r)^2+\Gamma^2}}\,,\\
\gamma''(r)&= \frac{\Gamma\gamma_0''}{\sqrt{\bar{\Omega}(r)^2+\Gamma^2}}\,,
\end{split}
\end{equation}
where $\gamma_0''$ is the imaginary part of the linear response
(\ref{linearresponse}),
\[\gamma_0'' = \frac{4e^2}{C}\omega|
\bar{\mathcal{A}}_0|^2\nu(\omega)f_{z}^0(\omega/2)\,,\] and bars indicate the
values of the functions at the resonance.  With this expression we find the
maximum response amplitude,
\begin{equation}
r_\text{m}=\tilde{I}_e\left(1-(\tilde{I}_e/{I}^\ast)^2\right)^{-1/2},
\end{equation}
where
\begin{equation}\label{}
\tilde{I}_e ={eI_e\over C\omega_p\gamma_0''}
={\omega_p\over 2\gamma_0''}\;{I_e\over I_C}
\end{equation}
is the dimensionless driving current,  and
\begin{equation}\label{}
{I}^\ast={\Gamma\over \omega_p \,|\bar{\mathcal{A}}_0|}\,.
\end{equation}
This result  shows that the response has an explosive instability manifested
by a divergency of the oscillation amplitude when the driving current amplitude
reaches the critical value $\tilde{I}_e=I^\ast$. We emphasize that this current is
much smaller than the Josephson critical current, $I_C$, which sets the scale
for the nonlinear behavior of the adiabatic junctions. This instability is
easy to understand noticing that the damping produced by the Andreev states
decreases with amplitude of oscillation, and, on the other hand, it is the
damping value that limits the resonance response amplitude. To eliminate the
divergency, one has to take into account other damping mechanisms, which are
weaker than the linear damping by the Andreev states.

If we turn off the external drive, $I_e=0$, we find from \Eq{slowdynamics}
the equation for the decay of the oscillation amplitude,
$\dot{r}=-\gamma''(r)r/2$. For $r > \Gamma/|\mathcal{\bar A}_0|\omega_p$, we
find that the plasma oscillation decays {\em linearly} with time with the
rate, $\dot{r}\approx -\left(\Gamma\gamma_0''/|\mathcal{\bar
A}_0|\omega_p\right)= $ const, until it enters the linear regime, $r <
\Gamma/|\mathcal{\bar A}_0|\omega_p$, where the decay crosses over to an
exponential time dependence, $r\sim \exp(-\gamma''_0 t)$.

\section{Langevin equation}
The classical equation of motion (\ref{classicaleqm}) is deterministic and
thus does not include the fluctuations originating from the coupling of the
phase to the quasiparticles. In this section we shall outline how these
effects can be taken into account. The method we adopt results in a quantum
Langevin equation \cite{Schmid,Weiss,Kamenev2005}, although as we shall show,
the stochastic force in this case generally has non-gaussian properties.

Expectation values of any function of phase, $F(\varphi)$, is given by
\begin{equation}
\langle F\rangle(t_1)=\int d\varphi_1 F(\varphi_1)\rho_\text{red}(\varphi_1,
\varphi_1,t_1),
\end{equation}
where $\rho_\text{red}(\varphi_1, \varphi_1,t_1)$ is the reduced density
matrix. Noticing that the partition function (\ref{Z}) is given by the trace
over the reduced density matrix, $Z=\int d\varphi_1
\rho_\text{red}(\varphi_1,\varphi_1,t)$, we are able to write the diagonal
elements in terms of the Wigner variables $\varphi,\chi$,  on the form,
\begin{equation}
\rho_\text{red}(\varphi_1,\varphi_1,t_1)=\mspace{-10mu}
\int_{\varphi(t_1)=\varphi_1}\mspace{-10mu}
\mathcal{D}\varphi\int_{\chi(t_1)=0}\mspace{-10mu}\mathcal{D}\chi
e^{iS[\varphi,\chi]},
\end{equation}
where the limits on the functional integrals indicate that the endpoints,
$\varphi(t_1),\chi(t_1)$, of the trajectories are to be held fixed.

To zeroth order in the saddle point approximation, $S[\varphi,\chi]\approx
\int dt \chi(t)\left(\delta S[\varphi,0]/\delta\chi(t)\right)$, only the
classical path, $\varphi_c(t)$, is realized and the density matrix is
written:
\begin{equation}\begin{split}
&\rho_\text{red}(\varphi_1,\varphi_1,t)\\&=\int_{\varphi(t_1) = \varphi_1}
\mathcal{D}\varphi\int_{\chi(t_1)=0} \mathcal{D}\chi \exp\left(i\int dt
\chi(t) \frac{\delta S[\varphi,0]}{\delta\chi(t)}\right)\\
&=\mspace{-10mu}\int_{\varphi(t_1)=\varphi_1} \mathcal{D}\varphi  \
\delta\left[\frac{\delta S[\varphi,0]}{\delta\chi(t)}\right],
\end{split}
\end{equation}
where, $\delta[\ldots]$, denotes a delta functional. Average quantities
are then entirely determined by the classical path $\langle F\rangle(t_1)
=F(\varphi_c(t_1))$.

To go beyond this deterministic description and include fluctuations we can
expand the action around the saddle point, $\chi=0$,  to second order,
\begin{equation}\begin{split}
S[\varphi,\chi]&\approx \int dt \chi(t)\frac{\delta
S[\varphi,0]}{\delta\chi(t)} \\
&+\frac{1}{2}\int dtdt'\chi(t)\frac{\delta^2
S[\varphi,0]}{\delta\chi(t)\delta\chi(t')}\chi(t').
\end{split}
\end{equation}
Here the kernel,
\begin{equation}\begin{split}
&i\frac{\delta^2S[\varphi,0]}{\delta\chi(t)\delta\chi(t')}\\&
=-\frac{i}{(2e)^2} \sum_{ab}\text{Tr}\left[\hat{G}^{ab}(t,t')
\hat{I}(t')\hat{G}^{ba}(t',t)
\hat{I}(t)\right]\\
&=-\frac{1}{(2e)^2} \mathcal{S}_I[\varphi](t,t'),
\end{split}
\end{equation}
is given by the symmetrized current-current correlation function,
$\mathcal{S}_I[\varphi](t,t')$, which is a functional of $\varphi$ due to
the dependence of $\hat{G}^{ab}$ and $\hat{I}$ on $\varphi(t)$.

We decouple the quadratic term in $\chi$ by introducing an auxiliary variable
$I_\xi$ which shall later be interpreted as a stochastic current
\cite{Schmid},
\begin{equation}\begin{split}
&\exp\left(-\frac{1}{2}\int dtdt'\chi(t)
\frac{1}{(2e)^2}\mathcal{S}_I[\varphi](t,t')
\chi(t')\right)\\
&=\int \mathcal{D}I_\xi e^{-\frac{i}{2e}\int dt I_\xi(t)\chi(t)}
P[\varphi,I_\xi],
\end{split}
\end{equation}
where $P[\varphi,I_\xi]$ denotes the functional distribution
\begin{equation}
P[\varphi,I_\xi]= \mathcal{N}[\varphi]e^{-\frac{1}{2}\int dtdt'I_\xi(t)\mathcal{S}_I^{-1}
[\varphi](t,t')I_\xi(t')},
\end{equation}
where $\mathcal{N}[\varphi]=(\text{det}S_I^{-1}[\varphi])^{-1/2}$.
The density matrix can then be written as
\begin{equation}\begin{split}
&\rho_\text{red}(\varphi_1,\varphi_1,t_1)\\
& =\int \mathcal{D}\xi \int \mathcal{D}\varphi \  \delta
\left[\frac{\delta
S[\varphi,0]}{\delta\chi(t)}-I_\xi(t)\right]P[\varphi,I_\xi].
\end{split}
\end{equation}
The delta functional selects a single trajectory, $\varphi_\xi$, for each
realization of, $I_\xi$, determined by the classical equation,
\begin{equation}\label{quantumlangevin}
\frac{\delta S[\varphi,0]}{\delta \chi(t)}=I_\xi(t) \Rightarrow
\frac{C}{2e}\ddot{\varphi}_\xi+\langle I \rangle[\varphi_\xi]=I_\xi\,,
\end{equation}
where, for the sake of convenience, we assumed an unbiased junction. Eq.
(\ref{quantumlangevin}), is a stochastic equation and averages are given
by,
\begin{equation}
\langle F\rangle (t_1)=\langle F(\varphi_\xi(t_1))\rangle_\xi,
\end{equation}
where $\langle \ldots \rangle_\xi=\int \mathcal{D}I_\xi (\ldots)
P[\varphi_\xi,I_\xi]$. In contrast to the conventional theory of quantum Langevin
equations the functional distribution, $P[\varphi_\xi,I_\xi]$, is in general
non-gaussian due to the dependence of the symmetrized current correlation
function on $\varphi_\xi=\varphi[I_\xi]$. This is a consequence of
non-equilibrium nature of the fermionic bath strongly coupled to the phase
variable.

The stochastic force becomes gaussian under the linear response
approximation. We consider small deviations from a classical equilibrium
configuration, $\varphi(t)=\varphi_0$ and $\hat{f}(t)=f^0$, and get the
equation,
\begin{equation}\label{Langevin}
\left(-\omega^2+\omega_p^2+\omega\gamma_0(\omega)\right)
\delta\varphi(\omega)=\frac{2e}{C}I_\xi(\omega).
\end{equation}
The functional distribution can be taken at the equilibrium value,
$P[\varphi_0,\xi]$, which then becomes Gaussian and the stochastic
current, $I_\xi$, satisfies the typical relations for Gaussian noise:
\begin{equation}
\langle I_\xi(t)\rangle_\xi=0, \qquad \langle I_\xi(t)I_\xi(t')
\rangle_\xi=\mathcal{S}^0_I(t-t'),
\end{equation}
 where
\begin{eqnarray}
&&\mathcal{S}^0_{I}(\omega) =  \coth \left(\frac{\omega}{2T}\right)
\sum_{ij}|I_{ij}|^2
(f_{i}^0-f_{j}^0)\pi\delta(\omega-\varepsilon_{ij})\nonumber\\
\displaystyle &=&\displaystyle
(2e)^2\omega^2\coth\left(\frac{\omega}{2T}\right)\sum_{ij}|
\mathcal{A}_{ij}|^2(f_{i}^0-f_{j}^0)\pi\delta(\omega-\varepsilon_{ij})\nonumber\\
&=&\displaystyle C\omega\coth\left(\frac{\omega}{2T}\right)
\text{Im}\gamma_0(\omega).
\end{eqnarray}
Thus the fluctuating current is related to the dissipative response by the
quantum fluctuation dissipation theorem.

\section{Conclusions}

We have presented a general  theory framework for describing non-adiabatic
dynamics of Josephson junctions with low energy quasiparticle states. The
theory applies to a wide class of Josephson junctions including transparent
mesoscopic contacts based on 2DEG, nanowires, quantum dots, and also
junctions of unconventional superconductors. It was shown that in the
classical limit the equation of motion for the phase must be solved together
with a Liouville equation for density matrix of low energy fermionic states.
Furthermore, we illustrated how the dynamics of such systems can differ
significantly from the adiabatic (tunnel) junctions, by investigating the
resonant dynamics of the phase and low energy Andreev bound states. It was
shown that nonlinear, two-state dynamics of the Andreev bound states, rather
than an adiabatic Josephson energy, defines the nonlinear macroscopic
dynamics of the junction.\\

This work was supported by the Swedish Research Council (VR), and the
European FP7-ICT Project MIDAS.

\newpage



\begin{thebibliography}{21}

\bibitem{Takayanagi}
H. Takayanagi  and T. Kawakami,  {\it Phys. Rev. Lett.} {\bf 54}, 2449
(1985).

\bibitem{Ruitenbeek}
N. van der Post, {\em et al}., {\it Phys. Rev. Lett.} {\bf 73} 2611 (1994).

\bibitem{Delft}
P. Jarillo-Herrero, J.A. van Dam, and L.P. Kouwenhoven, {\it Nature}
{\bf439}, 953 (2006).


\bibitem{Ryazanov}
V. V. Ryazanov, {\em et al}.,  {\it  Phys. Rev. Lett.} {\bf86}, 2427 (2001).

\bibitem{SNS}
I.O. Kulik, {\it  Zh. Eksp. Teor. Fiz.} {\bf 30}, 944 (1969) [Sov. Phys. JETP
{\bf 57}, 1745 (1969)].

\bibitem{ScS}
I.O. Kulik and A.N. Omel'yanchuk, {\it  Zh. Eksp. Teor. Fiz. Pis. Red.} {\bf
21}, 216 (1975) [JETP Lett. {\bf21} 96].

\bibitem{SFS}
I.O. Kulik, {\it Zh. Eksp. Teor. Fiz.} {\bf 49}, 1211 (1965) [Sov. Phys. JETP
{\bf 22}, 841 (1966)].

\bibitem{CB}
I.O. Kulik   and R.I. Shekhter,  {\it Zh. Eksp. Teor. Fiz.} {\bf68}, 623
(1975) [ Sov. Phys. JETP {\bf41}, 308 (1975)].

\bibitem{Nakamura1997}
Y. Nakamura, C.D. Chen, and J.S. Tsai, {\it Phys. Rev. Lett.} {\bf 79},
2328-2331 (1997).

\bibitem{Saclay1998}
V. Bouchiat, D. Vion, P. Joyez, D. Esteve, and M. H. Devoret,  {\it Phys.
Scr.} {\bf T76}, 165 (1998).

\bibitem{Friedman2000}
J. R. Friedman, V. Patel, W. Chen, S. K. Tolpygo. and J. E. Lukens, {\it
Nature} {\bf 406}, 43-46 (2000).

\bibitem{Leggett1985}
A. J. Leggett,  and A. J. Garg,  {\it Phys. Rev. Lett.} {\bf 54}, 857 (1985).

\bibitem{Makhlin2001}
Y. Makhlin, G. Sch\"{o}n and A. Shnirman  {\it Rev. Mod. Phys.} {\bf 73}, 357
(2001)

\bibitem{Wendin2007}
G. Wendin, and V. S. Shumeiko, {\it Low Temp. Phys.} {\bf 33}, 724 (2007).

\bibitem{Clark2008}
J. Clarke, F. K. Wilhelm,  {\it Nature} {\bf 453}, 1031 (2008).

\bibitem{Girvin2008}
R. J. Schoelkopf and S. M.  Girvin,  {\it Nature} {\bf 451}, 664 (2008).

\bibitem{Josephson}
B. D. Josephson,  {\it Rev. Mod. Phys.} {\bf 36}, 216 (1964).

\bibitem{Likharev}
K.K. Likharev, {\it Dynamics of Josephson Junctions and Circuits} (Gordon,
1986).

\bibitem{Scalapino}
D.J. Scalapino,   in {\em Tunneling Phenomena in Solids}, ed. E. Burstein and
S. Lundqvist (Plenum, 1969).

\bibitem{Beenakker1991}
C. W. J. Beenakker and H. van Houten, {\it Phys. Rev. Lett.} {\bf 66}, 3056
(1991).

\bibitem{Super96}
G. Wendin and V.S. Shumeiko, {\it Superlatt. and Microstr.} {\bf 20}, 569
(1996).

\bibitem{Lofwander2001}
T. L\"{o}fwander, V.S. Shumeiko,  and G. Wendin,   {\it Supercond. Sci.
Technol.} {\bf 14}, R53 (2001).

\bibitem{Scalapino1995}
D.J. Scalapino,  {\it Phys. Rep.} {\bf 250}, 329 (1995).

\bibitem{Eberly}
L. Allen  and J.H. Eberly,  {\it Optical Resonance and Two-Level Atoms},
(Dover, 1987).


\bibitem{AES1}
V. Ambegaokar, U. Eckern and G. Sch\"{o}n  {\it Phys. Rev. Lett.} {\bf 48},
1745 (1982).

\bibitem{Kamenev2005}
A. Kamenev,  arXiv:cond-mat/04122296v2, (2005).

\bibitem{Kleinert}
H. Kleinert, {\em Path Integrals in Quantum Mechanics, Statistics, Polymer
Physics, and Financial Markets} (World Scientific, 1999).

\bibitem{Weiss}
U. Weiss, {\em Quantum Dissipative Systems} (World Scientific, 2004).

\bibitem{ZAZ2}
A. Zazunov, V.S. Shumeiko, E.N. Bratus', and G. Wendin, {\it Phys. Rev. B}
{\bf 71}, 214505 (2005).

\bibitem{AES2}
U. Eckern, G. Sch\"{o}n and V. Ambegaokar {\it Phys. Rev. B} {\bf 30}, 6419
(1984).

\bibitem{LO}
A. I. Larkin and Yu. N. Ovchinnikov {\it Phys. Rev. B} {\bf 28}, 6281 (1983).

\bibitem{ZS}
A. D. Zaikin and G. Sch\"{o}n {\it Phys. Rep.} {\bf 198}, 237 (1990).

\bibitem{ZP}
A. D. Zaikin and S. V. Panyukov {\it Zh. Eksp. Teor. Fiz.} {\bf 89}, 242
(1985) [ {\it Sov. Phys. JETP} {\bf 62}, 137 (1985)]

\bibitem{CUEV}
A. Martin-Rodero, F. J. Garcia-Vidal and A. Levy Yeyati {\it Phys. Rev.
Lett.} {\bf 72}, 554 (1994)

\bibitem{FOG} J. C. Cuevas and M. Fogelstr\"{o}m {\it Phys. Rev. B}
 {\bf 64}, 104502 (2001).

\bibitem{ZAZ} A. Zazunov, V. S. Shumeiko, E. N. Bratus', J. Lantz, and
G. Wendin {\it Phys. Rev. Lett.} {\bf 90}, 0870003 (2003).

\bibitem{Furusaki}
A. Furusaki and M. Tsukada, {\it Physica B} {\bf 165-166}, 967 (1990).

\bibitem{PRB93}
V.S. Shumeiko, G. Wendin, and E.N. Bratus', {\it Phys. Rev. B} {\bf 48},
13129 (1993).

\bibitem{Schmid}
A. Schmid, {\em J. Low Temp. Phys.}  {\bf 49}, 609 (1982).


\end{thebibliography}
\end{document}